\documentclass[preprint,12pt,authoryear]{elsarticle}

\usepackage{amssymb}
\usepackage{amsmath}

\usepackage{lineno}
\usepackage{multirow}
\usepackage{threeparttable}
\usepackage{multirow}
\usepackage[ruled,vlined]{algorithm2e}
\usepackage{algorithmic}
\usepackage{subcaption}
\usepackage{graphicx}
\usepackage{enumitem}
\usepackage{booktabs}
\usepackage{amsthm}
\usepackage{stmaryrd}
\usepackage{bbding}
\usepackage{bbm}
\usepackage{caption}
\newcommand{\modelname}{RAU\xspace}
\usepackage{caption}   
\usepackage{graphicx}  
\usepackage{caption}
\usepackage{makecell}
\newcommand{\vpara}[1]{\vspace{0.2cm}\noindent\textbf{#1 }}

\begin{document}

\begin{frontmatter}
\title{RAU: Towards Regularized Alignment and Uniformity for Representation Learning in Recommendation}
\author[label1]{Xi Wu\fnref{eqauth}}
\ead{wuxi@stumail.ysu.edu.cn}

\author[label2]{Dan Zhang\fnref{eqauth}}
\ead{zd18@tsinghua.org.cn}

\author[label1]{Chao Zhou}
\ead{chaozhou@stumail.ysu.edu.cn}

\author[label3]{Liangwei Yang}
\ead{lyang84@uic.edu}


\author[label1]{Tianyu Lin}
\ead{lintianyu@stumail.ysu.edu.cn}

\author[label1]{Jibing Gong\corref{cor1}}
\ead{gongjibing@163.com}
\fntext[eqauth]{Both authors contributed equally to this work.}
\cortext[cor1]{Corresponding author.}

\affiliation[label1]{organization={School of Information Science and Engineering},
            addressline={Yanshan University}, 
            city={Qinghuangdao},
            postcode={066000}, 
            country={China}}
\affiliation[label2]{organization={Department of Computer Science and Technology},
            addressline={Tsinghua University}, 
            city={Beijing},
            postcode={100084}, 
            country={China}}
\affiliation[label3]{addressline={University of Illinois at Chicago}, 
            city={Chicago},
            country={USA}}
\begin{abstract}
Recommender systems (RecSys) have become essential in modern society, driving user engagement and satisfaction across diverse online platforms. Most RecSys focuses on designing a powerful encoder to embed users and items into high-dimensional vector representation space, with loss functions optimizing their representation distributions.
Recent studies reveal that directly optimizing key properties of the representation distribution, such as \textbf{\textit{alignment}} and \textbf{\textit{uniformity}}, can outperform complex encoder designs. However, existing methods for optimizing critical attributes overlook the impact of dataset sparsity on the model: limited user-item interactions lead to sparse alignment, while excessive interactions result in uneven uniformity, both of which degrade performance.

In this paper, we identify the sparse alignment and uneven uniformity issues, and further propose \underline{R}egularized \underline{A}lignment and \underline{U}niformity (RAU) to cope with these two issues accordingly.
RAU consists of two novel regularization methods for alignment and uniformity to learn better user/item representation. 
1)\textbf{ Center-strengthened alignment} further aligns the average in-batch user/item representation to provide an enhanced alignment signal and further minimize the disparity between user and item representation.
2) \textbf{Low-variance-guided uniformity} minimizes the variance of pairwise distances along with uniformity, which provides extra guidance to a more stabilized uniformity increase during training.
We conducted extensive experiments on three real-world datasets, and the proposed RAU resulted in significant performance improvements compared to current state-of-the-art CF methods, which confirms the advantages of the two proposed regularization methods.
\end{abstract}



\end{frontmatter}

\section{Introduction}

In the context of the increasing volume of online data~\citep{tang2008arnetminer, mayer2013big, zhang2024oag}, users can easily be overwhelmed by the massive information, resulting in the problem of information overload. Recommender Systems (RecSys) play a crucial role in helping users effortlessly identify items that match their interests, and thus have become an important part of modern society. These systems are integral to numerous online services, deeply integrating into our daily digital interactions, including but not limited to, news aggregation~\citep{wu2022news}, video game recommendations~\citep{yang2022large}, and e-commerce~\citep{lin2019cross}. 

Collaborative filtering (CF) is a pivotal technique in RecSys, leveraging historical user-item interactions to learn user's potential preferences. The most recent CF-based methods~\citep{IntegrateCF,LightGCN, zhang2023apegnn, DirectAU, GraphAU, zhang2024recdcl}focus on how to design a powerful encoder that encodes users and items into a high-dimensional space. These methods aim to convert historical user-item interactions into dense vector representations to learn meaningful embeddings for both users and items. These vectors reflect the user preferences and item characteristics, and their quality heavily affects the performance of RecSys. 

Generally, CF-based methods consist of three parts: Encoder, Negative Sampler, and Loss Function. Researchers have studied different parts to obtain a better user/item representation. The encoder is responsible for transforming raw user-item interaction data into meaningful representations. It captures the underlying patterns and characteristics in the data. Sophisticated encoders have been designed to generate more informative representations from the high-order interactions~\citep{LightGCN} and side information~\citep{yang2022large}. A Negative Sampler is crucial in distinguishing between items that a user has interacted with and those they have not. It generates negative samples (items not interacted with by the user) to provide contrastive data for the model. Negative samplers~\citep{wu2023dimension,ding2020simplify,ABNS} are also specifically designed to provide hard negative signals to enhance the model training. The study of loss functions for RecSys has long been overlooked until the finding of two critical desired properties~\citep{DirectAU} for user/item representation, \textit{i.e.}, alignment and uniformity. 

Alignment aims to align the interacted user and item representation together in the same direction on the normalized embedding space, while uniformity targets uniformly distributing the representations. The study~\citep{DirectAU} shows that even a direct optimization of alignment and uniformity on the embedding table can surpass the sophisticated designed encoders. Besides, it also saves sampling time as a negative sample-free method. It reveals the dominating role of loss function among the three parts in CF-based methods and rekindles our interest in deeply studying these two properties in this paper.

However, two critical problems exist with optimizing the alignment and uniformity property for RecSys.
\textbf{1) \emph{Sparse alignment.}} Data sparsity~\citep{GraphAU} is a common and crucial characteristic of RecSys data. Each user has only a limited amount of energy to interact with a few items, so the alignment signal is naturally sparse to train user/item representations. At present, there are two ways (GraphAU~\citep{GraphAU} \& MAWU~\citep{MAWU}) to enhance the alignment signals. GraphAU introduced a graph-based alignment to capture multi-hop relationships, but it is restricted to graph-based encoders. MAWU proposed a soft-border-based alignment that utilizes a learnable boundary to enhance alignment signals, but its effectiveness is insufficient because of the inherent data sparsity in recommendation systems.
\textbf{2) \emph{Uneven uniformity.}}
The ideal uniformity aims to maximize the sum of distances within user representations or item representations. When a distance is increased between the total user/item pairs, it may result in an unusually large distance between a few user/item pairs, while the distance between some user/item pairs may be small. However, this distribution does not distinguish users/items well. Therefore, relying solely on the distance between user pairs to determine uniformity is insufficient and may lead to uneven uniformity. To address these two critical challenges, we conducted detailed experiments and analysis, as presented in Section~\ref{Sec: AU Problems} where we quantify these challenges and provide insights into the limitations of existing methods.
\begin{table}[]
\centering
\caption{Critical comparison between existing AU-based models for recommendation.}
\begin{tabular}{c|c|c}
\specialrule{.10em}{.4ex}{.65ex}
Models & Explicit alignment & Guideline for \\
 & strength info. &  uniformity\\ 
 \hline
GraphAU \cite{GraphAU} & High-Order &  \XSolidBrush \\ \hline
MAWU \cite{MAWU} & Soft-Align & \XSolidBrush\\ \hline
RAU & Center-Strengthen & Low-Variance \\
\bottomrule
\end{tabular}
\label{table: compar}
\end{table}
In this paper, we propose two novel Regularization methods over Alignment and Uniformity (RAU) to tackle the aforementioned issues separately. 
\textbf{1) \textit{Center-strengthened Alignment}.} It is a novel Regularized Alignment (RA) method to further enhance the alignment signal by pulling in the distance between the center of the user and the item within the batch. The Center-strengthened alignment provides extra alignment signals to pull user and item representations directly into the same representation space. It is more effective across datasets with varying sparsity levels and is applicable to all existing recommendation scenarios, including both graph-based and non-graph-based approaches.
\textbf{2) \textit{Low-variance guided Uniformity}.} It is a Regularized Uniformity (RU) from the variance of the distance between user/item within a batch. The variance acts as the guideline for uniformity updating, to ensure that the uniformity increases steadily and mitigates the effect of extreme values on the overall uniformity.

Compared with existing methods for enhancing alignment and uniformity as shown in Table~\ref{table: compar}, RAU proposes a novel Center-Strengthened alignment that can provide more alignment signals to align users and items to the same representation space effectively. Besides, RAU is also the only method that identifies the drawbacks of uniformity and proposes effective solutions to guide uniformity training based on the low-variance principle.

The main contributions of this paper can be summarized as follows:
\begin{itemize}
    \item Conceptually, we point out the sparse alignment/unstable uniformity issue and propose feasible solutions for the regularized training of RecSys.
    \item Center-Strengthened Alignment: This new method enhances alignment by focusing on the distance between the centers of user and item representations within a batch. It provides additional alignment signals and is more effective in sparse datasets as it does not depend solely on historical user-item interactions.
    \item Low-Variance Guided Uniformity: This approach uses the variance of distances within a batch as a guideline for updating uniformity. It aims to ensure a steady increase in uniformity and mitigate the impact of extreme values, addressing the issue of unstable uniformity in RecSys.
    \item Superior performance: Extensive experiments are conducted on three real-world datasets, demonstrating the effectiveness of the two proposed regularization methods and validating the superiority of \modelname.
\end{itemize}

The rest of this paper is organized as follows. 
Section~\ref{Sec: Pre} introduces the preliminaries of the AU-based methods.
Section~\ref{Sec: AU Problems} discusses in detail the two critical problems in AU.
Section~\ref{Sec: Method} and Section~\ref{Sec: Expers} present the RAU method and report the performance.
Section~\ref{Sec: Related} reviews related work. Finally, concludes the paper and introduces the future work in Section~\ref{Sec: Conclusion}.

\section{PRELIMINARIES}\label{Sec: Pre}

\subsection{Collaborative Filtering}
CF-based methods are a foundational predictive approach within recommendation systems (RecSys), harnessing historical user interactions 
to learn a high-dimensional dense vector representation for each user and item. It selects the top-K items with the highest similarity scores for each user as a result of a recommendation to infer potential user preferences, which is a common and effective method for recommender systems. Esteemed for its efficacy, CF has been underscored as a potent instrument in seminal works~\citep{SimpleX,wu2023dimension, zhang2023apegnn}. 
At its core, CF delineates sets of users $U$ and items $I$, alongside a historical interaction matrix $R$, where an entry $R_{u, i}$ is marked as 1 to signify an interaction between user $u$ and item $i$, and 0 otherwise.
Innovative CF methodologies~\citep{LightGCN, NGCF} have adopted the BPR loss function~\citep{BPR-MF} as a mechanism to refine encoder functions $f$(·). These functions are tasked with distilling user and item attributes into representative embedding vectors $\mathbf{e}_{u}, \mathbf{e}_{i} \in \mathcal{R}_{d}$, with $d$ embodying the dimensionality of the embedding space.
The similarity score $s(u, i)$, indicative of the predicted preference, is computed through measures of similarity—such as the dot product—between the embeddings of user $u$ and item $i$, formally denoted as $s(u, i) = e_{u}^{T}e_{i}$. This scoring underpins the recommender system's ranking process, where items are ordered by predicted relevance. The system then distills these rankings to extract the top-K items, forming a tailored recommendation list for each user. This technique, while straightforward, remains a quintessentially effective paradigm in RecSys, balancing simplicity with performance.
\subsection{Alignment and Uniformity}
\begin{table}[]
\caption{Summary of the main notations}
\label{tab: Notations}
\begin{tabular}{ll}
\hline
Notations & Descriptions \\ \hline
    $U \left ( u \in U,u{}' \in U \right ) $      &     User set         \\
    $I \left ( i \in I,i{}' \in I \right ) $      &     Item set         \\
    $\mathcal{R}$      &       The whole rating matrix      \\
    $\mathcal{R}_{u,i} $      &  The rating score of user $u$ to item $i$           \\
    $\mathcal{R}_{d} $      &    The dimension size of embeddings         \\
    $p_{data}$      &      The distribution of whole data        \\
    $p_{pos}$      &       The distribution of positive pairs data       \\
    $p_{user}, p_{item}$      &      The user/item distribution of positive pairs data        \\
   $d(x,y )$ &        The distance between two representations$\left ( e_{x},e_{y} \right ) $      \\
    $\mathbf{e}_{u}\in p_{pos},\mathbf{e}_{i} \in p_{pos}$      &    The user/item embedding of user $u$/item $i$          \\
    $\mathbb{E}$      &     Calculation of average value         \\ \hline
\end{tabular}
\end{table}
Recent literature~\citep{DirectAU, MAWU, GraphAU} has highlighted two principal characteristics pivotal for collaborative filtering algorithms' efficacy: alignment and uniformity. Alignment is the concept that similar data points should be mapped closely in the embedding space, thereby encouraging the embedding vectors of positive pairs to be nearer to each other. Uniformity, on the other hand, aims to ensure that the embeddings are spread uniformly across the embedding space, preventing the model from collapsing into a narrow set of points and instead encouraging a more informative distribution over the embedding hypersphere. These two principles are considered critical to the success of representation learning, with the dual objectives of clustering positive samples closely while ensuring that unrelated samples' embeddings are adequately dispersed. 

The key to the success of these AU-based methods is how to map high-dimensional dense vector representations onto the unit hypersphere and compute the distances of the representation distributions. 
Considering the whole data distribution (denoted as $p_{data}(\cdot)$) and the distribution of positive sample pairs (denoted as $p_{pos}(\cdot)$), $L_{2}$ regularization is applied to map these representations onto the unit hypersphere. The distance between two representations is represented as $d(x,y)$. For example, the distance between a user $u$ and the item $i$ he has interacted with can be calculated using Equation~\ref{eq: representation distance} below, and the alignment loss is directly calculated using the distance (Equation~\ref{eq: align}).
\begin{align}
d(u,i)=||\mathbf{e}_{u}-\mathbf{e}_{i} ||^{2}_{2},
\label{eq: representation distance}
\end{align}
\begin{align}
\mathcal{L}_{align}= \underset{(u,i)\in p_{pos}}{\mathbb{E}}  \left \{ d(u,i) \right \} ,
\label{eq: align}
\end{align}
\begin{align}
\mathcal{L}_{uniform}= \frac{1}{2} log\underset{(u,u{}')\in p_{data}}{\mathbb{E}}\left \{ e^{-2\cdot d(u,u{}')} \right \} +\frac{1}{2}log\underset{(i,i{}' )\in p_{data}}{\mathbb{E}}\left \{ e^{-2\cdot d(i,i{}')} \right \}  .
\label{eq: uniform}
\end{align}
where $\widetilde{\mathbf{e}}$ denotes the $l_{2}$ normalized embedding.
Uniform distribution on the unit hypersphere is considered to be the problem of minimizing the total pairwise potential under the action of a certain kernel function~\citep{wang2020understanding}. In recommender systems, one transforms the distances between like representations (between users internally or between itemsets internally) into a potential minimization problem using a Gaussian kernel function~\citep{DirectAU}. The main notations used in this study are listed in Table~\ref{tab: Notations}.

\section{Problems with alignment and uniformity}\label{Sec: AU Problems}
In this section, we will first empirically examine the effect of sparsity on alignment and evaluate the difference of alignment on convergence speed as well as recommendation performance. Subsequently, we provide a theoretical exploration of the uneven uniformity issue, revealing its implications for representation quality and distinguishability.

\subsection{Problems with Alignment} \label{sub:ra}
\begin{figure}[htbp]
    \centering 
    \includegraphics[width=0.7\textwidth]{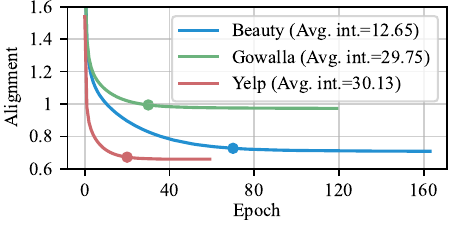}
    \caption{Alignment change along the training on different datasets with varied average user/item interaction numbers.}
    \label{fig: align}
\end{figure}
Previous CF-based research focused on designing powerful encoders to obtain more collaborative signals from training data, enhancing user and item representations, and then optimizing these representations using BPR loss.
In contrast, the AU-based approach maps the user and item representations onto a hypersphere, optimizing the distribution by directly improving two key properties: alignment and uniformity. 
However, this approach does not address the challenges posed by data sparsity in recommender systems.
Recently, some studies have noted the sparsity issue in alignment optimization, and attempts have been made to alleviate this problem by introducing additional alignment-enhancing signals. Typical designs include:
\begin{itemize}[leftmargin=*,itemsep=0pt,parsep=0.5em,topsep=0.3em,partopsep=0.3em]
\item \textbf{Graph-based alignment:} e.g. GraphAU~\citep{GraphAU}, which introduces the graph structure to alleviate the lack of alignment signals by incorporating multi-hop neighbor alignment signals between users and items. However, introducing graph structures means it can only work with graph-based encoders, which contradicts the original intent of AU-based methods. The original purpose of AU-based methods was to improve recommendation performance by directly optimizing desired properties in collaborative filtering through new learning objectives, rather than designing complex encoders.
\item \textbf{Soft-border based alignment:} e.g. MAWU~\citep{MAWU}, which assigns a learnable boundary to enhance the alignment signal for user-item pairs with interaction history. However, as mentioned earlier, existing recommender systems already face the challenge of data sparsity. Enhancing alignment signals by learning a boundary is far from insufficient in this context.
\end{itemize}

While both GraphAU and MAWU are powerful tools, they still struggle to address the issue of sparse alignment signals fully.
This inspired us to design an encoder-independent sparse alignment enhancement method that can quickly adapt to various scenarios with different sparsities.

To better understand how alignment affects recommendation performance, we selected three recommendation scenarios with varying levels of sparsity to train the DirectAU model, the first AU-based model proposed for collaborative filtering. To ensure fairness between users and items, we measured the sparsity of each scenario using the average number of interactions per user and per item, with more specific data provided in Table~\ref{tab:dataset}. MF was used as the backbone for training DirectAU, and we recorded the changes in alignment during training across the three scenarios, as shown in Figure~\ref{fig: align}. 
Additionally, to explore the differences between AU-based methods and traditional BPR-based optimization, we also trained a MF encoder with the BPR loss function. The performance improvement of DirectAU compared to the BPR-equipped MF encoder is shown in Table~\ref{tab:Improvement}.

\begin{table}
\centering 
\caption{Improvement of DirectAU over BPR-MF}
\label{tab:Improvement}
\begin{tabular}{ccccccc} 
\toprule 
           & \multicolumn{2}{c}{Gowalla} & \multicolumn{2}{c}{Beauty} & \multicolumn{2}{c}{Yelp} \\
\cmidrule(lr){2-3} \cmidrule(lr){4-5} \cmidrule(lr){6-7} 
           & R@20 & N@20 & R@20 & N@20 & R@20 & N@20 \\
\midrule 
BPR-MF     & 13.88     & 8.22    & 10.51    & 4.85    & 6.85    & 4.22 \\
DirectAU   & 20.01     & 11.66   & 14.07    & 6.77    & 11.03   & 6.87 \\
\textbf{Improv(\%)} & \textbf{44.2} & \textbf{41.8} & \textbf{33.9} & \textbf{39.6} & \textbf{61.0} & \textbf{62.8} \\
\bottomrule 
\end{tabular}
\end{table}


It is common knowledge that the average number of interactions per user/item in different recommendation scenarios is different, naturally leading to differences in alignment signals available for model training.  
As shown in Table~\ref{tab:dataset}, the average number of interactions for Yelp and Gowalla are similar and significantly higher than those for Beauty. Observing Figure~\ref{fig: align}, we can find that the alignment signals for Yelp and Gowalla converge at similar speeds, much faster than for Beauty. The convergence speed ranking of alignment signals across these three scenarios aligns perfectly with their average interaction counts. This insight prompts new thinking about alignment optimization: users and items receive richer alignment signals in scenarios with higher average interaction counts. Conversely, alignment signals are sparser with fewer interactions, requiring more training steps. 
Moreover, the improvement brought by the AU-based approach of using key attributes to optimize the model compared to the traditional optimization using BPR, which emphasizes ranking, also corresponds to the average number of interactions.
The data in Table~\ref{tab:Improvement} indicates that richer alignment signals correspond to lower alignment values for the model, and the performance improvements brought by the AU-based method are greater compared to traditional methods.
The AU-based method achieves the most significant improvements with richer alignment signals, such as Yelp and Gowalla, with an improvement of up to 62.8\% on Yelp. 
These findings inspire us to explore more effective and adaptive methods for enhancing alignment signals.
\subsection{Problems with Uniformity}
\begin{figure*}[htbp]
  \centering
  \begin{subfigure}[b]{0.4\linewidth}    \includegraphics[width=0.8\linewidth]{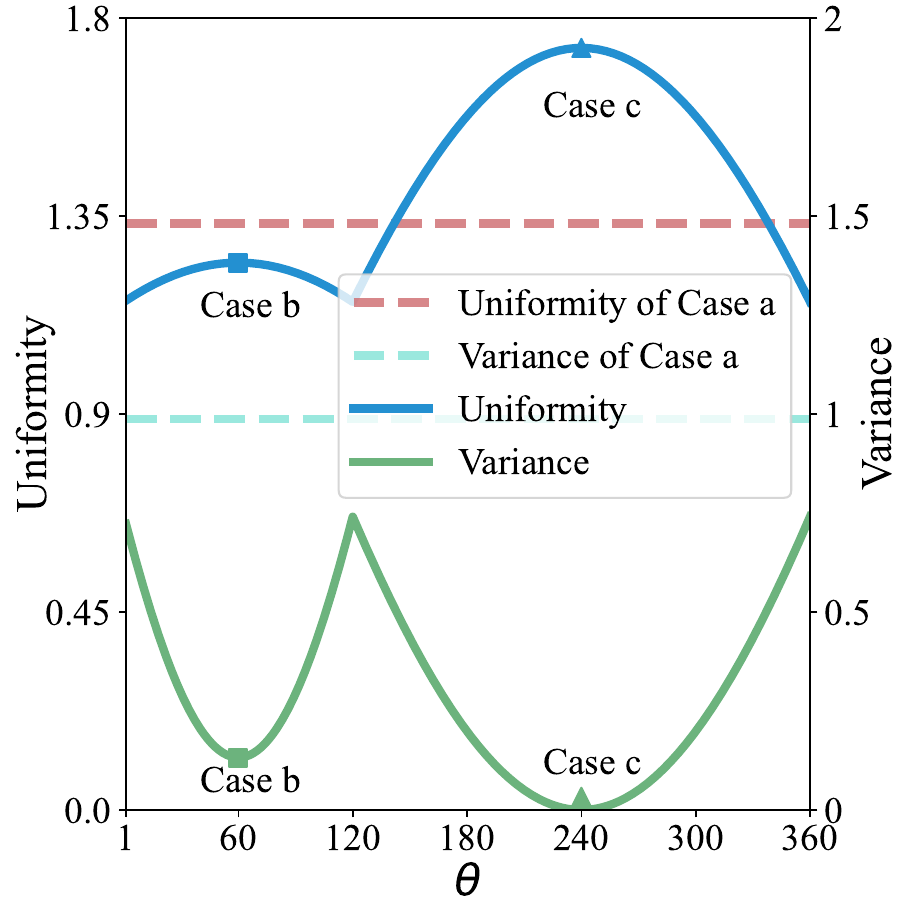}
    \caption{}
    \label{fig: uniform_move}
  \end{subfigure}
  \centering
  \begin{subfigure}[b]{0.55\linewidth}
    \includegraphics[width=\linewidth]{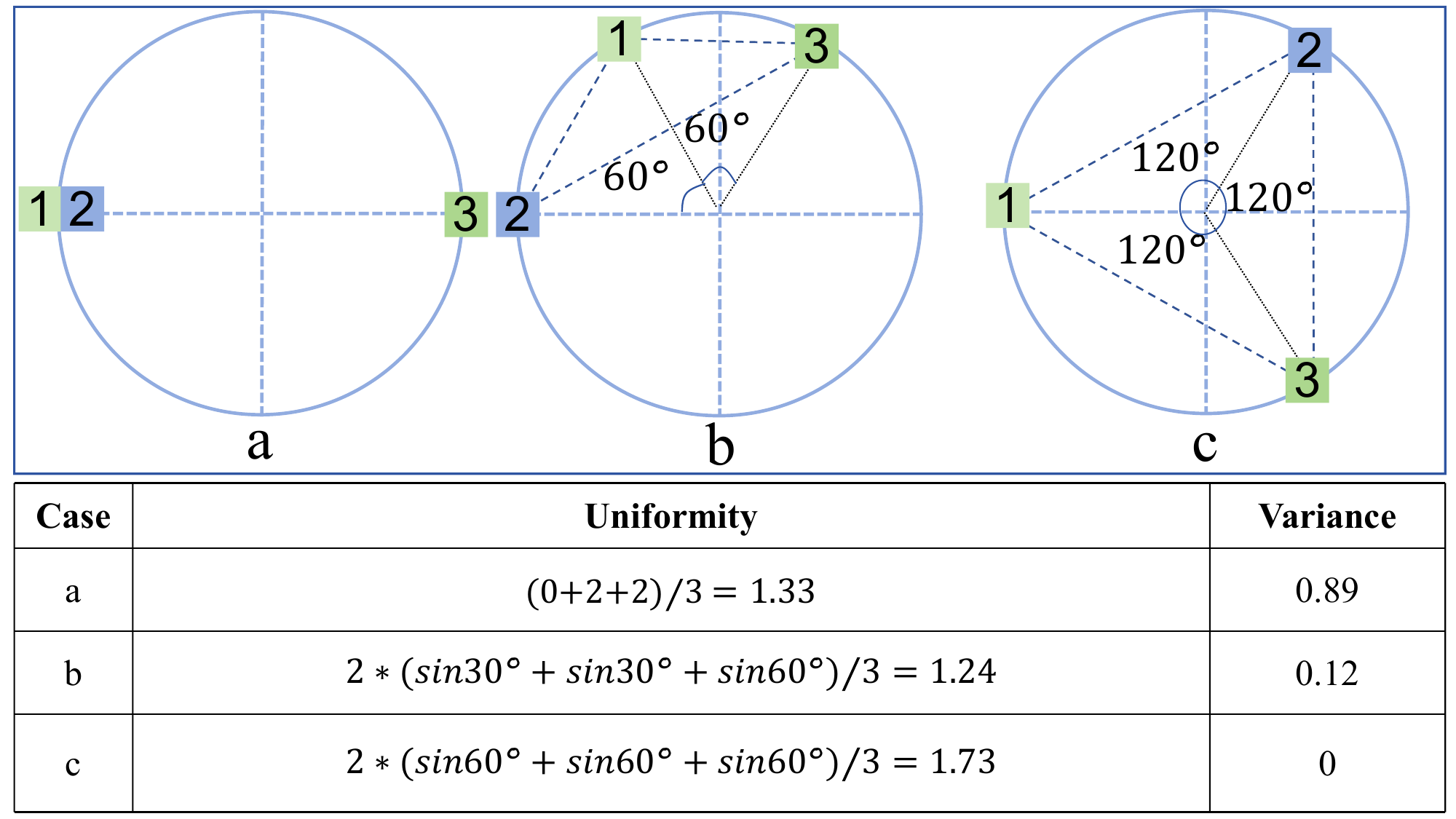}
  \caption{}
  \label{fig: uniform_point}
  \end{subfigure}
  \caption{The sample points are distributed on the unit circle. We find that randomly increasing uniformity does not necessarily achieve uniformity.}
  \label{fig: uniform}
\end{figure*}
After experimentally investigating the impact of sparsity on alignment, we further have found through theoretical analysis that sparsity has the same impact on uniformity optimization.
As mentioned in the uniformity calculation Equation~\ref{eq: uniform}, the uniformity calculation problem on the hypersphere is transformed into the potential minimization problem between two representations based on a kernel function. The key to this problem lies in the distributional distance~\ref{eq: representation distance} between the two representations.
\begin{align}
f_{Gaussian}(\cdot ) = \mathbb{E}_{\left \{ \left ( x,y\right )\in \left \{ U \, or\, I \right \}   \right \}}   \sum_{\left ( x,y\right )  }e^{-2d(x,y )}
\label{eq: potential minimization}
\end{align}
As shown in Equation~\ref{eq: potential minimization}, to achieve potential minimization on the hypersphere, it is essential to maximize the distance of the representations distributed above.

However, in recommendation scenarios with strong alignment signals, the distances between users and items tend to become excessively close. Traditional uniformity optimization methods, which only focus on users or items, often lead to representation overlap, resulting in uneven issues, as illustrated in case (a) of Figure~\ref{fig: uniform_point}. Moreover, the likelihood of such overlap increases sharply with stronger alignment signals. Ideal uniform optimization should resemble case (c) in Figure~\ref{fig: uniform_point}, where the distances between the three points are maximized. We conducted interesting experiments to explore strategies for minimizing representation overlap and achieving the ideal uniform distribution.
We began by analyzing the uniformity of three points located at different positions on the unit circle and measured the variance between the three points.
Specifically, Figure ~\ref{fig: uniform_move} demonstrates the changes in uniformity and variance of three points based on the movement of a point. Figure ~\ref{fig: uniform_point} exhibits three distinct distributions: case a is the extreme case, with the poorest uniformity; case b features a moving point forming an angle of 60 degrees with one of the points, providing improved uniformity; case c represents the best uniformity of the ideal distribution.
From Figure ~\ref{fig: uniform_point} we find that the value of uniformity for case a is greater than the value for case b. Therefore, case a's uniformity is better than case b's according to DirectAU's definition. However, the actual situation is that case a has the worst uniformity. We find from Figure ~\ref{fig: uniform_move} that the value of uniformity when the angle between two points is 120 degrees and the other point is a moving point has about half the probability of being smaller than the value of uniformity for case a. 
From the above analysis, we find that randomly increasing the uniformity between samples does not achieve a truly uniform distribution.
Another interesting phenomenon is that the variance of the three points is always less than the variance of the case no matter what position the moving point is in, showing that the uniformity increases the variance decreases and the variance takes the minimum value when the uniformity takes the maximum value.

Therefore, we chose to use the variance of the distance between samples as a guideline for updating the model uniformity. In other words, our goal is to steadily increase the uniformity between samples to achieve a Low-variance uniformity.
\label{sub: ru}
\section{The Regularized Alignment and Uniformity}\label{Sec: Method}
\begin{figure}[htbp]
    \centering 
    \includegraphics[width=0.9\textwidth]{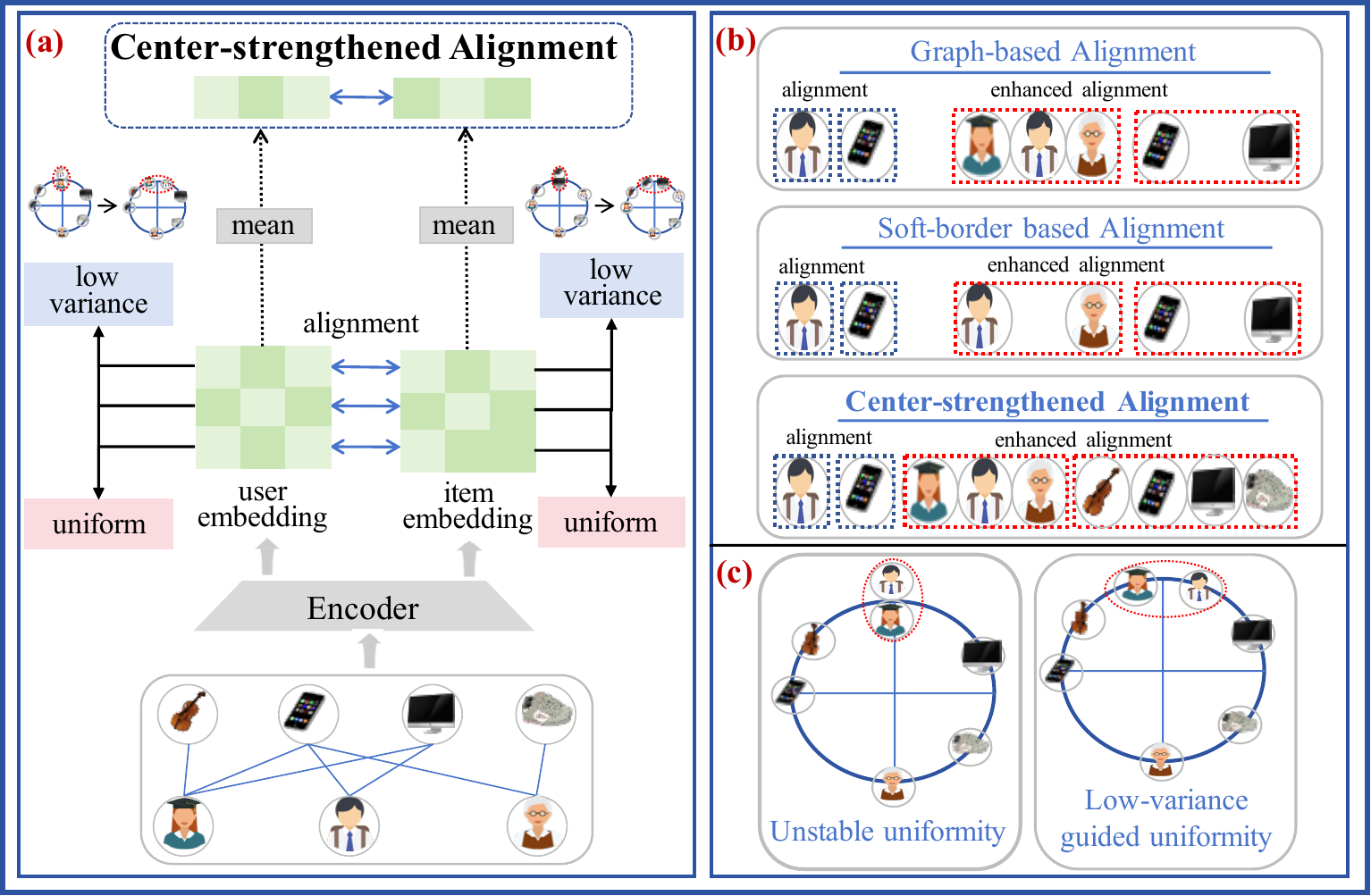}
    \caption{Overview of the \modelname framework.}
    \label{fig: framework}
\end{figure}
The preceding analysis highlights the importance of enhancing alignment signals in sparse datasets and guiding uniformity to increase progressively for effective learning of user/item representations.
To address these problems, We introduce two innovative \textbf{R}egularization methods for \textbf{A}lignment and \textbf{U}niformity (RAU), specifically designed to tackle the issues above individually.
\textbf{1) \textit{Center-strengthened Alignment}} is a new method called \textbf{R}egularized \textbf{a}lignment (RA) that aims to improve the alignment signal by bringing the center of the user and the center of the item closer together within the batch. The Center-strengthened Alignment provides additional alignment signals to help bring user and item representations into the same representation space. This method is particularly effective for sparse datasets as it doesn't rely heavily on sparse historical user-item interactions.
\textbf{2) \textit{Low-variance guided uniformity}} is a method that helps ensure uniformity in a batch by using the variance of the distance between users and items. This method is called \textbf{R}egularized \textbf{U}niformity (RU). Using the variance as a guideline for updating uniformity, ensures that the uniformity consistently increases and helps to reduce the impact of extreme values on overall uniformity.
Figure \ref{fig: framework} illustrates the overall structure of the proposed framework.
The flow chart of the model is shown in \ref{fig: framework}(a). The user and item are passed through the Encoder to obtain the user and item embeddings. Firstly, the user-item interaction pairs are aligned, and the alignment signal is enhanced using center-strengthened alignment. Then, the Low-variance strategy is used to guide the update of the user/item uniformity within the batch to prevent unstable uniformity. Figure \ref{fig: framework}(b) describes three strategies for alignment enhancement. Figure \ref{fig: framework}(c) describes the difference between unstable uniformity and Low-variance guided uniformity.
\subsection{Center-strengthened Alignment}
To further fortify the alignment signal, we have implemented an alignment regularization technique. This method is particularly designed to intensify the attraction between positive sample pairs. Concurrently, it ensures a consistent and cohesive pull-in effect on the entire distribution of users and items within each batch. This dual strategy not only addresses the shortfall in alignment signals but also empowers the model to assimilate alignment information more proficiently and comprehensively. The specific formulation for the alignment regularization loss is outlined below:
\begin{align}
\mathcal{L}_{RA} &= ||\mathop{E}\limits_{(u, i)\in p_{pos}}(\widetilde{\mathbf{e}}_u - \widetilde{\mathbf{e}}_{i})||^2
\label{eq:ralign}
\end{align}
\subsection{Low-variance Guided Uniformity}
We develop a uniformity regularization strategy (RU), aimed at enhancing the uniformity within our model. This strategy utilizes the variance of the user/item distance within a batch to direct a smooth and steady increase in uniformity. The definition of RU is as follows:
\begin{align}
\mathcal{L}_{RU} =  \mathop{E}\limits_{(u,u^\prime)\in p_{user}}(e^{-2\cdot d(u,u^\prime)}- \mathop{E}\limits_{(u,u^\prime)\in p_{user}}e^{-2\cdot d(u,u^\prime)})\\
+\mathop{E}\limits_{(i,i^\prime)\in p_{item}}(e^{-2\cdot d(i,i^\prime)}- \mathop{E}\limits_{(i,i^\prime)\in p_{item}}e^{-2\cdot d(i,i^\prime)}).
\label{eq:runiform}
\end{align}
We achieve better uniformity by regularizing the uniformity, given the direction in which the uniformity increases, i.e., letting the distance between the samples increase while also letting the variance of the distance between the samples decrease as much as possible, and letting the model smoothly increase the uniformity.
\subsection{Objective}
The MAWU method \citep{MAWU} posits that applying uniform weights for user and item uniformity is not feasible across diverse datasets, due to variations in their inherent diversity. Consequently, MAWU recommends customizing weights for each user and item following their individual Gini coefficients. A higher Gini coefficient \citep{chin2022datasets} signifies increased inequality, necessitating a greater variance and weight in the distribution.
\begin{algorithm}[h]
  \caption{The training process with \modelname}
  \label{algo}
  \renewcommand{\algorithmicrequire}{\textbf{Input:}}
  \renewcommand{\algorithmicensure}{\textbf{Output:}}
  \begin{algorithmic}[1]
  \REQUIRE Training set $\mathcal{G}=\{(u,i^{+})\mid u \in \mathcal{U}, i^{+}\in \mathcal{I}  \}$, Recommendation user/item encoder $f_\theta(\cdot)$, 
  \ENSURE User/Item representations $\left \{ e_{u}\mid u \in U \right \} $ and $\left \{ e_{i}\mid i \in I \right \} $
  \STATE Initialize $\left \{ e_{u}\mid u \in U \right \} $ and $\left \{ e_{i}\mid i \in I \right \} $
  \FOR{$t= 1,2,3, \dots ,$ to $T$}
  \STATE Sample a mini-batch of user-item positive pairs$\{(u,i)\}$.
  \STATE Initialize loss $\mathcal{L}_{RAU}=0$.
  \FOR{each mini-batch $\{(u,i)\}$ pair}
  \STATE Get the user and positive item embeddings $\{(\mathbf{e}_u, \mathbf{e}_i)\}$ by encoder $f_\theta(\cdot)$.
  \STATE Calculate Align loss by Eq.~\eqref{eq: align};
  \STATE Calculate RA loss by Eq.~\eqref{eq:ralign};
  \STATE Calculate WU loss by Eq.~\eqref{eq:wu};
  \STATE Calculate RU loss by Eq.~\eqref{eq:runiform};
  \STATE Calculate $\mathcal{L}_{RAU}$ by (\ref{eq:rau}).
  \ENDFOR
  \STATE Update representations by descending the gradients
    \ENDFOR
  \end{algorithmic}
\end{algorithm}
Adopting this rationale, we have implemented a weighted uniformity loss approach. Our empirical findings affirm the efficacy of this method. The formula for weighted uniformity loss is defined as follows:
\begin{align}
\mathcal{L}_{WU}= \gamma_{1}log\underset{(u,u^{'})\in p_{user}}{E}e^{-2\cdot d(u,u^{\prime})}\\
+\gamma_{2}log\underset{(i,i^{'})\in p_{item}}{E}e^{-2\cdot d(i,i^{\prime})}.
\label{eq:wu}
\end{align}
To address the issues associated with DirectAU, we introduce a novel loss function called \modelname.
This loss function incorporates both RA and RU.
\begin{align}
\mathcal{L}_{RAU} = \mathcal{L}_{align}+\mathcal{L}_{WU}+\alpha \mathcal{L}_{RA}+\beta \mathcal{L}_{RU}.
\label{eq:rau}
\end{align}
\subsection{Complexity Analysis}
Considering the efficiency of \modelname, we present the training process step by step in Algorithm~\ref{algo} and provide its runtime in Table~\ref{tab:runtime}.
For a fair comparison, the models in Table~\ref{tab:runtime} are all implemented in the same framework, using a 2-layer GCN as the encoder.
As observed from Algorithm~\ref{algo}, the computational time complexity of each training epoch for RAU, compared to other AU-based methods, differs from the newly introduced computation of two regularization signals.
The computational complexity of GraphAU is $\mathcal{O}\left ( 2KBd+2B^{2}d \right ) $, but the nested and indexed representations across multiple layers on the graph result in a longer runtime for each step. MAWU introduces an additional learnable parameter and modifies the uniformity calculation, leading to a computational complexity of $\mathcal{O}\left ( 4Bd+2B^{3}d \right ) $. In contrast, RAU only utilizes a single-layer representation to calculate the enhancement signals, resulting in a computational complexity of $\mathcal{O}\left ( Bd+4B^{2}d \right ) $.

It is worth noting that while RAU introduces some manageable computational complexity, it offers broader applicability, being suitable for both graph-based and non-graph-based recommendation scenarios. Moreover, it delivers significant performance improvements, making the trade-off highly worthwhile.
\begin{table}[]
\centering
\caption{Runtime statistics on Gowalla dataset (seconds).}
\label{tab:runtime}
\begin{tabular}{lccc}
\hline
Method   & \multicolumn{1}{l}{time/epoch} & \multicolumn{1}{l}{\#epoch} & \multicolumn{1}{l}{total time} \\ \hline
GraphAU  &  60.5     &  114     &   6895.7          \\
MAWU     &  158    &    34      &    5386   \\
LightGCN &  205   &   180    & 36936       \\ \hline
RAU      & 27     &   39     &   1053   \\ \hline
\end{tabular}
\end{table}
\section{EXPERIMENTS}\label{Sec: Expers}
We empirically evaluate the proposed RAU on three real datasets. The aim is to answer the following five research questions (RQs).
\begin{itemize}[leftmargin=*,itemsep=0pt,parsep=0.5em,topsep=0.3em,partopsep=0.3em]
    \item \textbf{RQ 1:} Does \modelname boost recommendations with regularization? 
    \item \textbf{RQ 2:} Is RAU still working on the Graph-based backbone?
    \item \textbf{RQ 3:} How do these two regularization methods work?
    \item \textbf{RQ 4:}  Can RAU produce stronger alignment signals as well as achieve better uniformity?
    \item \textbf{RQ 5:} What is the impact of different hyper-parameters on RAU?
\end{itemize}

\subsection{Experimental Setup}
\subsubsection{Datasets}
We utilized three publicly available datasets of different sizes and sparsities. Specific details for each dataset are described in Table ~\ref{tab:dataset}. We split the dataset into training, validation, and test sets following the same process as DirectAU~\citep{DirectAU}. 
Consequently, we split the interactions of each user into training, validation, and test sets at a ratio of 0.8/0.1/0.1.
\begin{itemize}[leftmargin=*,itemsep=0pt,parsep=0.5em,topsep=0.3em,partopsep=0.3em] 
    \item \textbf{Beauty} \footnote{https://jmcauley.ucsd.edu/data/amazon/links.html} consists of a series of product reviews on Amazon.
    \item \textbf{Gowalla}~\footnote{http://snap.stanford.edu/data/loc-gowalla.html} is a dataset of user check-in locations from location-based social networking sites. 
    \item \textbf{Yelp} \footnote{https://www.yelp.com/dataset} is a business domain recommendation dataset including user reviews of products.
\end{itemize}
\begin{table}[htbp]
\centering
\caption{Statistics of the public datasets.}
\begin{tabular}{c|ccccc}
\toprule
Dataset  & \#Users & \#Items & \#Inter & \#Avg. int.& \#Avg. int. \\
  &  &  &  & per user & per item \\ \hline 
Beauty & 22.4K  & 12.1K   & 198.5K & 8.9 &16.4  \\ 
Gowalla  & 29.9K & 41.0K   &   1027.4K & 34.4 & 25.1 \\
Yelp2018   & 31.7K & 38.0K   & 1561.4K    & 49.3 & 41.1\\ 
\bottomrule
\end{tabular}
\label{tab:dataset}
\end{table}
\subsubsection{Evaluation Metrics}
We use two standard evaluation metrics, Recall@K and Normalized Discounted Cumulative Gain (NDCG@K), to evaluate the effectiveness of our model in making recommendations in recommender systems. These metrics are applied to measure the quality of the top-k recommended items by our model.
In our evaluation, we adhere to the DirectAU setting and report the results for two specific values of K, namely K=20 and K=50. To conduct the evaluation, we employ a full-ranking technique where we rank all items that the user has not yet interacted with. This technique is based on prior research. We then calculate the average metrics across all users in the test set.
\subsubsection{Baseline Models}
To assess the effectiveness of our \modelname, we compared it with several other relevant models, including the early (POP), MF-based models (NeuMF and BPR-MF), GNNs-based models (LightGCN and NGCF), SSL-based models (SGL, BUIR, CLRec), and AU-based models (MAWU and DirectAU). Below is a detailed description of our comparison.
\label{app:baselines}
\begin{itemize}[leftmargin=*,itemsep=0pt,parsep=0.5em,topsep=0.3em,partopsep=0.3em]
    \item Pop curates recommendations by selecting items that have historically garnered significant popularity and acclaim.
    \item BPR-MF \citep{BPR-MF} optimizes MF using BPR loss by implementing negative sampling on user-item interactions.
    \item NeuMF \citep{NeuMF} model conceptualizes users and items based on implicit feedback, utilizing a multi-layer perceptron for this process.
    \item NGCF \citep{NGCF} introduces a message-passing framework, utilizing both first-order and high-order propagation methods to execute graph convolution for collaborative filtering.
    \item LightGCN \citep{LightGCN} eschews the typical GCN steps of feature transformation and non-linear activation. Instead, it learns representations by linearly propagating user and item embeddings on a user-item interaction graph.
    \item BUIR \citep{BUIR} forms representations of users and items by estimating specific statistics exclusively from repeated random sampling of existing data, rather than relying on negative samples.
    \item SGL~\citep{SGL} employs diverse data augmentation techniques within a graph structural context to facilitate contrastive learning. This approach not only hastens the convergence process but also significantly strengthens the robustness of feature learning.
    \item XSimGCL~\citep{XSimGCL} indeed proposes a recommendation algorithm based on graph contrastive learning, and designs a simple yet effective data augmentation technique that involves adding noise.
    \item CLRec \citep{CLRec} introduces a negative sample queue mechanism, ensuring each sample has the possibility of being handled as a negative sample, thereby mitigating exposure bias.
    \item DirectAU \citep{DirectAU}  deconstructs the BPR loss into two separate components: alignment and uniformity, and optimizes them independently. This strategy effectively reduces the distance between orthogonal pairs while maintaining the maximal informational content in each representation, leading to enriched user/item representations
    \item MAWU \citep{MAWU} introduces an innovative loss function that incorporates Margin-aware Alignment, which takes into account item popularity, along with Weighted Uniformity
\end{itemize}
\begin{table*}[htbp]
\centering
\caption{Comparison of overall top-20 and top-50 performances~(\% is omitted) with representative models on three datasets. The best and second-best results are in bold and \underline{underlined}.}
\vspace{-0.2cm}
\renewcommand{\arraystretch}{1}
\setlength{\tabcolsep}{1.0mm}{
\resizebox{\textwidth}{!}{ 
\begin{tabular}{c|c|c|c|c|c|c|c|c|c|c|c|c|c} 
\specialrule{.16em}{0pt}{.65ex}
\multirow{2}{*}{Models} & Dataset & \multicolumn{4}{c|}{Gowalla} & \multicolumn{4}{c|}{Beauty} & \multicolumn{4}{c}{Yelp}   \\ 
\cmidrule(lr){2-14}

& Metrics      & R@20 & R@50    & N@20& N@50& R@20 & R@50    & N@20& N@50& R@20 & R@50    & N@20& N@50\\ 
\specialrule{.10em}{.4ex}{.65ex}
Base & Pop       & 3.24 & 5.13 & 1.66  & 2.12 & 3.25 & 5.80&1.31 &1.82 & 1.59& 3.06&  0.96&1.37\\
\specialrule{.10em}{.4ex}{.65ex}
\multirow{2}{*}{MF-based}
& BPR-MF   &13.88 &22.12 &8.22 & 10.24& 10.51&16.29 &4.85& 6.03& 6.85&12.82 & 4.22& 5.59\\
&NeuMF       & 11.24 & 19.06 & 6.07 & 7.98 & 5.55& 9.81& 2.38&3.26 & 5.14&10.03 &3.11&4.53\\
\specialrule{.10em}{.4ex}{.65ex}
\multirow{2}{*}{GNNs-based} & NGCF       & 13.76  &  22.32 & 8.15 & 10.24  & 9.95& 15.59  & 4.48& 5.63 &6.57&   13.74 & 3.92& 6.22  \\
& LightGCN     &14.80  & 24.72  & 8.07 & 10.46 & 12.14&  18.11 & 5.82& 7.05  & 9.17&   16.87& 5.65&  7.87 \\

\specialrule{.10em}{.4ex}{.65ex}
 \multirow{5}{*}{}& BUIR         & 11.80 & 19.13 & 6.85 & 8.62  & 11.21&  17.86 & 5.10&   6.43& 7.16&13.45   & 4.35& 6.17 \\
& SGL\_ED     &  18.88  & 29.63  & 11.22  & 13.85  &  12.84 & 19.35  & 6.04  & 7.38  & 9.33  & 16.94  & 5.84  & 8.04  \\ 
SSL-based& SGL\_ND     &  19.14  & 29.94  &  11.38 &  14.02 & 12.73  & 19.11  &  6.05 & 7.36  &9.23   & 16.71  & 5.77  & 7.94  \\ 
& SGL\_RW     &  18.76  & 29.39  & 11.04  & 13.64  & 12.98  & 19.70  &  6.16 & 7.53  & 9.20  & 16.86  &  5.57 &  7.97 \\ 
& CLRec     &  17.82  & 28.47  & 10.48  & 13.07  & 13.17  &  19.42 & 6.54 &  7.83 & 10.57& 18.59 &6.62& 8.94 \\ 
& XSimGCL &20.62 & 30.94&17.40  &21.07 &13.29 & 19.56& 6.37&7.66 &10.90 &19.44 & 6.90&9.37\\
\specialrule{.10em}{.4ex}{.65ex}
\multirow{4}{*}{AU-based}& DirectAU  & 20.01 &31.33&	11.66&	14.43 &14.07 &\underline{20.68} &6.77& 8.23 &{11.03} &\underline{19.32} &{6.87}&{9.26}\\ 
& MAWU &\underline{20.20}& \underline{31.52} &\underline{11.79} & \underline{14.56} &\underline{14.13} &{20.27}&\underline{6.99}&\underline{8.25}&\underline{11.05}&{19.27}&\underline{6.91}&\underline{9.28}  \\
&\textbf{\modelname}   &\textbf{20.98} &\textbf{32.56} &\textbf{12.40} &\textbf{15.22} & \textbf{14.58} & \textbf{21.50} & \textbf{7.06} & \textbf{8.47} & \textbf{11.36}& \textbf{19.76}& \textbf{7.11}& \textbf{9.55} \\
\cmidrule(lr){2-14} 
&\textbf{Imp (\%)}   &3.86&3.30&	5.17&	4.53&3.18&3.97&	1.00&2.67&2.81&2.54&	2.89&2.91 \\
\specialrule{.16em}{.4ex}{0pt}
\end{tabular}
}}
\label{tab: all_results}
\end{table*}
\subsubsection{Implementation Details}
\label{ssec:hyper}
All experiments were implemented based on the RecBole \citep{zhao2021recbole} framework for a fair comparison.
Specifically, we use Xavier ~\citep{glorot2010understanding} to initialize the embedding parameters and use the Adam optimizer~\citep{kingma2014adam} with a learning rate of \{$1e^{-3}$\} for all methods.
The batch size for training is set to 256 on the Beauty dataset, and 1024 on the Gowalla and Yelp datasets. In order to maintain fairness, the embedding sizes for all models are fixed at 64. Additionally, an early stop mechanism is triggered when the NDCG@20 score drops for 10 consecutive epochs on the validation dataset.
In \modelname, the default encoder is typical matrix factorization (MF) between users and items.
We tune the coefficient $\beta$ with $\mathcal{L}_{RU}$ in range of \{1$\sim$15\} and $\alpha$ with $\mathcal{L}_{RA}$ in range of \{0.0$\sim$1.0\}. The weight $\frac{\gamma_{1}}{\gamma_{2}}$ of $\frac{user}{item}$ in of $\mathcal{L}_{WU}$ are tuned in the range of $\{\frac{0.5}{0.5}, \frac{0.6}{0.4}, \frac{0.7}{0.3}, \frac{0.8}{0.2}, \frac{0.9}{0.1}\}$. 
For all baselines, we carefully followed the hyperparameters set in the respective original papers.
\subsection{RQ1: Overall Performance On MF}
Table \ref{tab: all_results} presents the effectiveness of the four distinct benchmark CF approaches as well as our suggested \modelname. The findings of our experiment enable us to make several crucial insights. Intriguingly, RAU yields substantial performance improvements, unusual given that the majority of the baseline methods were created from the recent two years of research, including the most recent AU-based. Here are some key observations derived from the outcomes:

\vpara{Comparison with MF-based models.} 
The GNN-based model outperforms the MF-based model across all Yelp datasets. However, BPR-MF achieves better results than NGCF on Beauty and Gowalla, suggesting that BPR loss maintains its superiority. Both the SSL-based model and the AU-based model outperform the MF-based model on every dataset, indicating a superior ability to learn robust representations for users/items.

\vpara{Comparison with GNNs-based models.} Table \ref{tab: all_results} shows that, in addition to BUIR, AU-based models outperform GNN-based models in terms of recall and NDCG on all datasets.
Our proposed model, named \modelname, has achieved remarkable results. It has surpassed the state-of-the-art GNNs-based model, LightGCN, by a significant margin of 41.76\%, 23.90\%, and 20.10\% in terms of Recall@20 on three different datasets, namely Gowalla, Yelp, and Beauty. This result demonstrates that acquiring an effective representation is more important in enhancing performance than creating a sophisticated encoder.

\vpara{Comparison with SSL-based models.} 
Table \ref{tab: all_results} shows that the AU-based model outperforms the SSL-based model in all three datasets. The key to this enhanced performance lies in the AU-based model's capacity to precisely adjust the balance between alignment and uniformity. This direct manipulation of these critical factors enables the model to achieve optimal performance across different datasets. Given the distinct characteristics inherent to each dataset, the AU-based model's capability to fine-tune this equilibrium is especially crucial.

\vpara{Comparison with AU-based models.} 
RAU outperformed the AU-based model across all datasets, showing the most significant improvement on the Gowalla dataset. This indicates that strengthening the alignment signal and employing low variance as a guiding principle for uniform performance can further optimize the effectiveness of AU-based models.
\subsection{RQ2: Performance on Graph-based Backbone.}

\begin{table*}[htbp]
\centering
\caption{Overall top-20 and top-50 performances~(\% is omitted) comparison with Different Models in Graph-based Backbone.}
\vspace{-0.2cm}
\renewcommand{\arraystretch}{1}
\setlength{\tabcolsep}{1.0mm}{
\resizebox{\textwidth}{!}{ 
\begin{threeparttable} 
\begin{tabular}{c|c|c|c|c|c|c|c|c|c|c|c|c|c} 
\specialrule{.16em}{0pt}{.65ex}
\multirow{2}{*}{Models} & Dataset & \multicolumn{4}{c|}{Beauty} & \multicolumn{4}{c|}{Gowalla} & \multicolumn{4}{c}{Yelp} \\    
\cmidrule(lr){2-14}
& Metrics      & R@20 & R@50    & N@20& N@50& R@20 & R@50    & N@20& N@50& R@20 & R@50    & N@20& N@50\\ 
\specialrule{.10em}{.4ex}{.65ex}
\multirow{4}{*}{\makecell{LightGCN \\ as \\ Backbone}} & DirectAU      & 14.49 &21.73 & 6.79 & 8.27 & 19.75&31.44 & 11.59&14.42 &11.24&19.67 &7.10&9.54\\ 
& MAWU  &14.72 &21.43 &7.21 &8.59 &20.62 &32.23 &12.03 &14.86&12.29 &19.85 &7.11 &9.58 \\
& GraphAU & 13.47 & 20.28& 6.51&7.90 &14.39 &21.16 &6.85 &8.25 &9.99 &17.81 &6.27 &8.53 \\
\cmidrule(lr){2-14}
\cmidrule(lr){2-14}
\cmidrule(lr){2-14}
&\textbf{\modelname}   &\textbf{14.86}  &\textbf{21.73} &\textbf{7.22} & \textbf{8.59}&\textbf{21.34} & \textbf{32.74}&\textbf{12.66} &\textbf{15.46} & \textbf{12.31} & \textbf{19.90} & \textbf{7.20}	& \textbf{9.60}
\\
\specialrule{.16em}{.4ex}{0pt}
\end{tabular}
\end{threeparttable} 
}}
\label{tab: results_lightgcn}
\end{table*}

We conducted a series of experiments to evaluate the effectiveness of \modelname when applied to the Graph-based backbone. We compare the effectiveness of the AU-based model on the current state-of-the-art graph-based backbone (LightGCN).
As depicted in Table \ref{tab: results_lightgcn}, not only does \modelname surpass the MF backbone, but so do DirectAU and MAWU. This further highlights that GNNs can mitigate the negative impact of data sparsity on recommendation performance. Additionally, RAU continues to achieve superior results on all three datasets within the LightGCN-based backbone, further reinforcing the effectiveness of RAU.
Overall, RAU performs well in the LightGCN-based backbone, especially on the Gowalla dataset, where it significantly outperforms the other two models.
\subsection{RQ3: Ablation Study}
We performed ablation experiments on our model, \modelname, across three distinct datasets: Beauty, Gowalla, and Yelp. In these studies, we systematically eliminated components associated with different regularization methods to evaluate their contributions.

As shown in the table \ref{table: rau ablation}, where $w/o_{-RA}$ denotes the removal of the Center-strengthened alignment module. $w/o_{-RU}$ denotes the removal of the uniformity module for Low-variance guided uniformity. Table \ref{table: rau ablation} presents the RAU ablation studies on Gowalla, Beauty, and Yelp. The ablation studies include the ablation of Center-strengthened alignment and the ablation of Low-variance guided uniformity. 
Our findings suggest that both Center-strengthened alignment and Low-variance guided uniformity are required for improving recommendation accuracy in AU-based models. Furthermore, we observed that removing Low-variance guided uniformity has a greater impact on recommendation accuracy than removing Center-strengthened alignment. This implies that achieving true uniformity is crucial for improving the performance of AU-based models.
In particular, the Yelp dataset exhibits a unique situation where the model is optimized when the coefficient of RA is 0, i.e., when the Low-variance guided uniformity module ($w/o_{-RU}$) is removed. This phenomenon suggests that the role of model components may vary for different datasets, and thus requires careful tuning and optimization for specific dataset sparsity in practical applications.
\begin{table}[htbp]
\centering
\caption{Ablation Study of RAU’s components and strategies on Gowalla and Beauty. }
\begin{tabular}{c|cc|cc|cc}
\specialrule{.10em}{.4ex}{.65ex}
\multirow{2}{*}{Models} & \multicolumn{2}{c|}{Gowalla} & \multicolumn{2}{c|}{Beauty} & \multicolumn{2}{c}{Yelp}\\ 
\cmidrule(lr){2-7}
 & R@20    & N@20   & R@20 & N@20  & R@20 & N@20       \\ 
 \hline
DirectAU & 20.01&  11.66& 14.07& 6.77 &11.03 &6.87\\ \hline
{$w/o_{-RA}$}& 20.83 & 12.40   & 14.37  &6.93 &11.36 &7.11           \\ 
{$w/o_{-RU}$} &  20.54 &   12.14  &  14.29 &6.80 &11.03 &6.87  \\ 
\hline
RAU & 20.98&  12.40& 14.58& 7.06 &11.36 &7.11\\
\bottomrule
\end{tabular}
\label{table: rau ablation}
\end{table}
\subsection{RQ4: Alignment and Uniformity Study}
In order to explore in depth the effect of RAU alignment strategies and their performance in enhancing uniformity, this study conducted experiments on the Beauty data set. We compare the effect of alignment under different strategies, including the no-enhancement strategy(DirectAU), the soft-boundary-based enhancement strategy(MAWU), and the center-enhancement strategy(RAU), and analyze the trend of these strategies with the training period as well as their effect on the recall@20 metric. In addition, on the Gowalla and Yelp datasets, we also examined the patterns of uniformity and variance as a function of training period.

\begin{figure}
  \centering
  \begin{subfigure}[]{0.4\linewidth}
    \includegraphics[width=\linewidth]{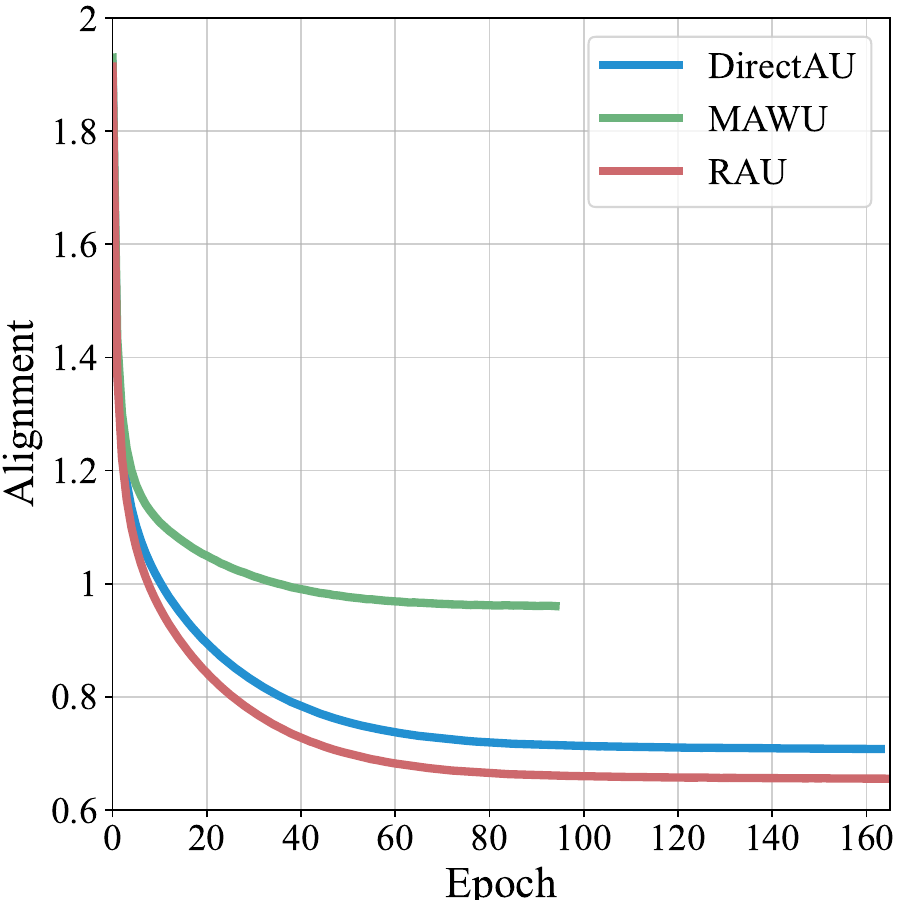}
    \captionsetup{skip=0pt}
    \caption{}
    \label{fig: Beauty_loss}
  \end{subfigure}
  \begin{subfigure}[]{0.4\linewidth}
    \includegraphics[width=\linewidth]{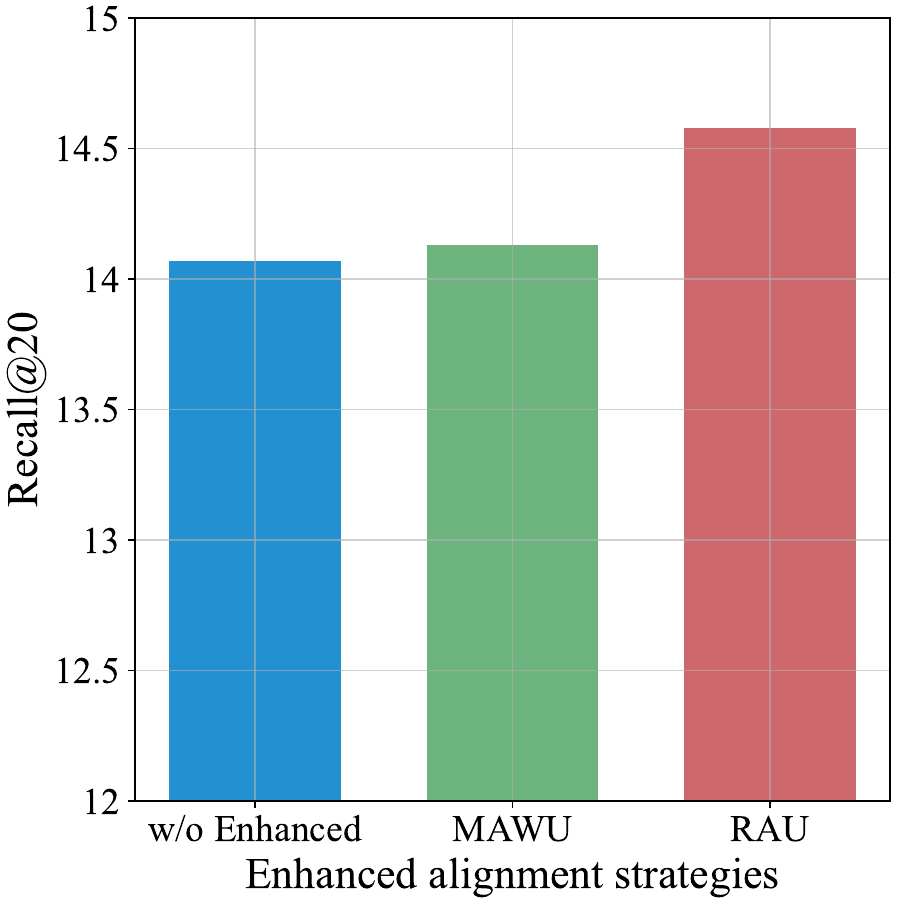}
    \captionsetup{skip=0pt}
    \caption{}
    \label{fig: Beauty_result}
  \end{subfigure}
  \caption{Performance of different alignment strengths.}
  \captionsetup{skip=-12pt}
  \label{fig:ra_effect}
\end{figure}

\vpara{Different Alignment Enhancements.} We implemented two distinct alignment enhancement strategies: MAWU and \modelname, on the sparsest Beauty dataset. The progressive change in alignment with each epoch for both methodologies was visualized in Figure \ref{fig:ra_effect}, along with the outcomes associated with Recall@20. 
Figure ~\ref{fig: Beauty_loss} illustrates that the MAWU converges rapidly, albeit with a substantial alignment loss. This suggests that while this strategy demonstrates robust alignment enhancement capability during the initial pre-training phase, it fails to maintain this effectiveness in the later stages when sample points have learned a better representation.
As a result, it ultimately leads to a higher alignment loss.
The strategy that uses RAU converges faster and achieves lower convergence values compared to the previous unenhanced strategy. This suggests that the Center-strengthened strategy is effective throughout the training project. 
Figure ~\ref{fig: Beauty_result} shows that on the Beauty dataset, the RAU Recall@20 outperforms both the MAWU and the no-alignment enhancement approach.

\begin{figure}[htbp]
    \centering \includegraphics[width=0.87\textwidth]{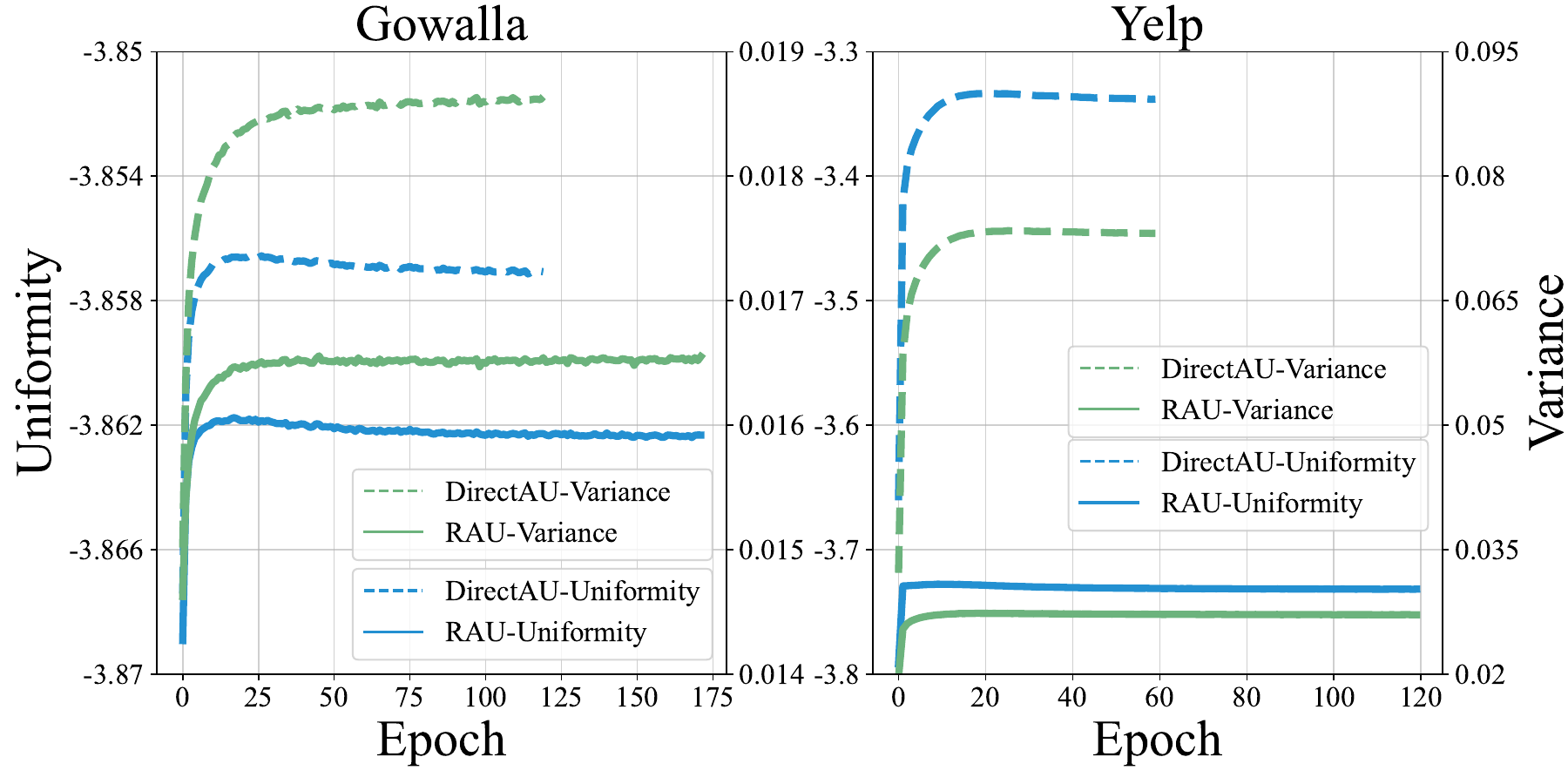}
    \caption{Comparative Analysis of Uniformity Loss and Variance Before and After RU Implementation.}
    \label{fig:ru_effect}
\end{figure}

\vpara{Can Low-variance guided uniformity be obtained with better uniformity?}
We compared the uniformity loss of DirectAU and RAU models on two datasets, Gowalla and Yelp. RAU model incorporates the low-variance guided strategy.
The analysis results are shown in Figure \ref{fig:ru_effect}. Our analysis reveals that RAU systematically generates a reduced uniformity loss when benchmarked against DirectAU on both the Gowalla and Yelp datasets. This suggests that RAU is superior in distributing users/items more uniformly across the recommendation space.
Moreover, RAU not only reduces uniformity loss across these datasets but also demonstrates a decreased variance in the distances measured between pairs of users/items. This indicates a more consistent recommendation spread, evidencing less variance in the users/items pairs distances. 
This stability is particularly advantageous as it suggests that RAU is not only distributing its recommendations more evenly but is also doing so consistently for different users.
Therefore, RAU effectively enhances uniformity by applying a low-variance uniformity approach.
\begin{figure*}
  \centering
  \begin{subfigure}[]{0.3\columnwidth}
\includegraphics[width=\textwidth]{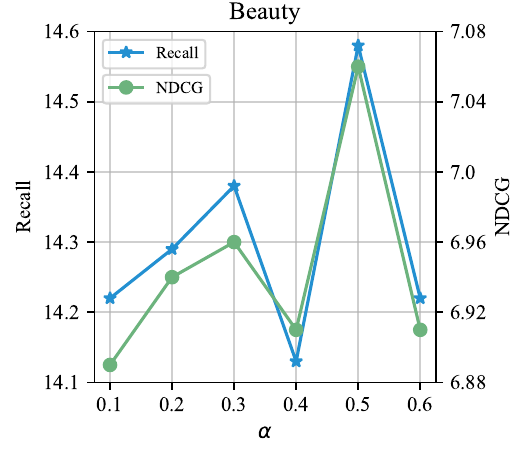}
    \caption{}
    \label{fig: alpha0}
  \end{subfigure}
  \begin{subfigure}[]{0.3\columnwidth}
\includegraphics[width=\textwidth]{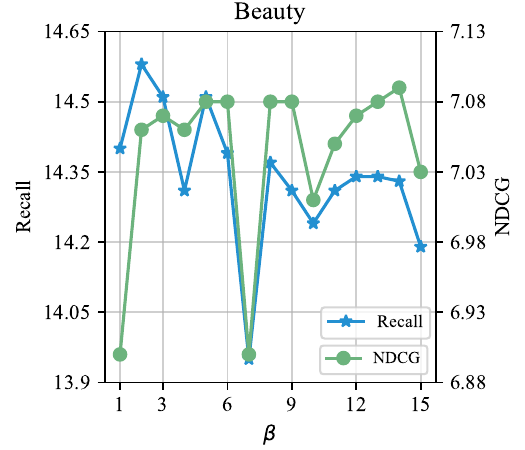}
    \caption{}
    \label{fig: beta0}
  \end{subfigure}
  \begin{subfigure}[]{0.3\columnwidth}
\includegraphics[width=\textwidth]{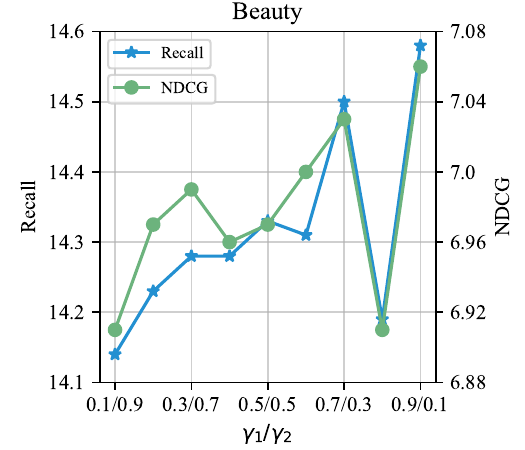}
    \caption{}
    \label{fig: gamma0}
  \end{subfigure}
  \begin{subfigure}[b]{0.3\columnwidth}
\includegraphics[width=\textwidth]{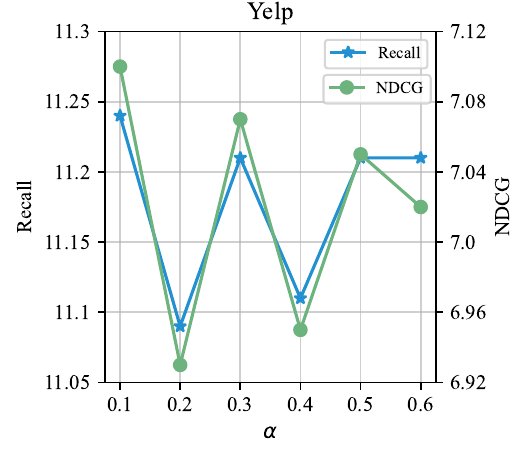}
    \caption{}
    \label{fig: alpha1}
  \end{subfigure}
  \begin{subfigure}[b]{0.3\columnwidth}
\includegraphics[width=\textwidth]{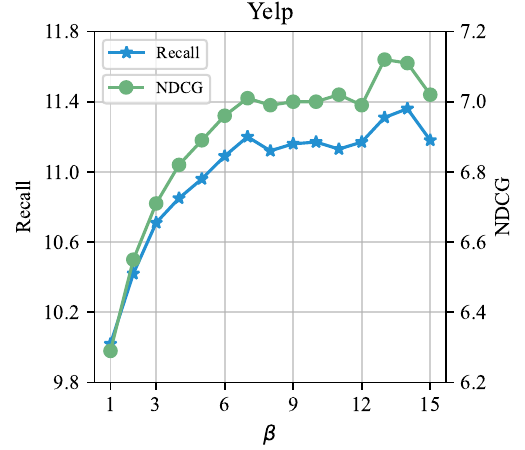}
    \caption{}
    \label{fig: beta1}
  \end{subfigure}
  \begin{subfigure}[b]{0.3\columnwidth}
\includegraphics[width=\textwidth]{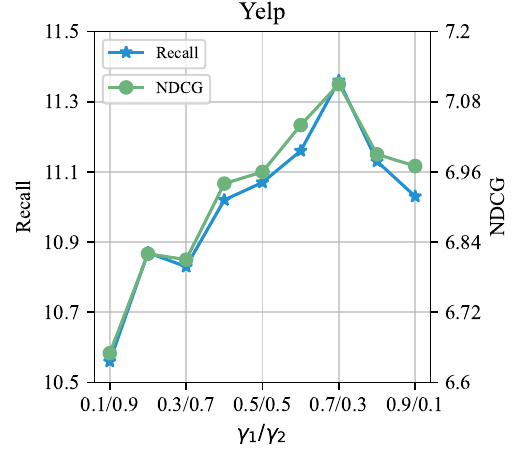}
    \caption{}
    \label{fig: gamma1}
  \end{subfigure}
  \caption{Hyper-Parameter Analysis on Beauty and Yelp.}
\end{figure*}
\subsection{RQ5: Parameter Sensitivity}
The main hyperparameters of RAU are 1) the weights of Center-strengthened alignment and Low-variance guided uniformity, and 2) the ratio of user uniformity to item uniformity.

\textbf{Evaluation of the Parameters $\alpha$ and $\beta$}
We further explored the influence of the parameters $\alpha$ and $\beta$ through a series of experimental analyses. The outcomes are depicted in Figures \ref{fig: alpha0} and \ref{fig: alpha1}. These figures reveal the varying impacts of $\alpha$ across different datasets, showcasing distinct probability distributions associated with this parameter. It is evident that as the dataset density increases, the optimal $\alpha$ value for peak performance decreases. Conversely, in datasets with higher sparsity, a larger $\alpha$ value is beneficial, reinforcing our earlier claim that enhancing signal strength is crucial in sparse data environments.

Additionally, the parameter $\beta$ is observed to be closely tied to the dataset density. In the sparse Beauty dataset, lower $\beta$ values (e.g., $\beta$=2) yield superior Recall@20 metrics. In contrast, for the dense Yelp dataset, higher $\beta$ values (e.g., $\beta$=14) result in the most favorable performance. As the density of the dataset increases, the propensity for extreme value distributions becomes more pronounced, necessitating a stronger uniformity regularization, and thus, a higher $\beta$ value. This observation aligns perfectly with our previous theoretical discussions.

\vpara{Impact of $\gamma_1$ and $\gamma_2$ Ratios.} 
%
Given that the user distribution offers unique insights compared to the item distribution \citep{MAWU}, we assign separate weights to user and item uniformity. Adopting this approach, we investigated the corresponding hyperparameters. Figures \ref{fig: gamma0} and \ref{fig: gamma1} delineate the impact of the $\gamma_1$ to $\gamma_2$ ratio on recommendation accuracy for the Beauty and Yelp datasets.
Our experimental findings yield several key observations:
(i) The performance distribution varies across different datasets. Specifically, the optimal $\gamma_1$ to $\gamma_2$ ratios for the Beauty and Yelp2018 datasets are 9 and 7/3, respectively. This suggests that when the item distribution is skewed or largely imbalanced, the focus on item uniformity should be lessened.
(ii) Across all datasets, consistently superior performance is achieved when $\gamma_1$ is greater than $\gamma_2$. This indicates that achieving uniformity in user representations is more critical than in item representations. Consequently, it is essential to fine-tune $\gamma_1$ and $\gamma_2$ in accordance with the dataset's unique characteristics. For instance, in the case of a skewed item distribution, the emphasis on item uniformity should be reduced.

These insights underscore the importance of calibrating the $\gamma_1$ and $\gamma_2$ weights to optimize recommendation system performance, taking into account the specific properties of the user and item distributions within each dataset.

\section{RELATED WORK }\label{Sec: Related}
\subsection{Collaborative Filtering}
Collaborative filtering (CF) is a technique used by some of the most successful recommendation systems to predict the preferences of a user by collecting preferences from many users (collaborating)~\citep{RecVAE, Mult-VAE, MacridVAE, BUIR, EGLN, MAWU, NCL, 10.1145/3442381.3449844, SimpleX, CF, SelfCF, yu2022self}. The underlying assumption of the CF approach is that those who agreed in the past tend to agree again in the future. 
Traditional research is mainly based on matrix factorization encoders to encode users and items separately~\citep{MF, NeuMF}. 
It utilizes a predefined loss function (common loss functions are BPR loss~\citep{BPR-MF}) to allow the encoder to learn the collaborative signals between users and items~\citep{BUIR, SGL, CLRec,ALHARBE2023119380}. 
With the development of graph neural networks in various fields, researchers have also introduced graph neural networks into recommendation systems and developed a series of graph-based collaborative filtering models (such as NGCF~\citep{NGCF}, LightGCN~\citep{LightGCN}, and so on). 
These collaborative filtering models based on graph neural networks take the historical interaction of users and items as a bipartite graph and utilize the connectivity of graph neural networks to capture higher-order collaborative filtering signals with remarkable success~\citep{wu2023dimension,zhang2023apegnn,WANG2025125605}.
\subsection{Alignment and Uniformity in CF}
There are a large number of recent research findings~\citep{gao-2021-simcse,wang2020understanding, DirectAU} in computer vision (CV), natural language processing (NLP), and Graph field showing that unsupervised learning can help improve the effectiveness of representation learning.
Simultaneously, recent studies~\citep{DirectAU, SimGCL, Wu_SSM, zhang2024recdcl} in the field of recommendation systems have indicated that significant enhancements in model recommendation performance can be achieved through the redefinition of appropriate loss functions. 
In particular, traditional matrix factorization (MF) models~\citep{BPR-MF} equipped with recommendation effects from certain self-supervised tasks have outperformed even state-of-the-art graph neural network (GNN)-based models~\citep{LightGCN,NGCF,zhang2023apegnn}. 
This has greatly inspired researchers on how to design more efficient loss functions.
Study~\citep{wang2020understanding} firstly proposes that a good representation can be learned by distinguishing two key properties on the hypersphere: Alignment and Uniformity.
DirectAU~\citep{DirectAU} designed a more effective collaborative filtering loss function by theoretically analyzing the connection between the traditional BPR-loss~\citep{BPR-MF} and these two properties in recommender systems.
Further, MAWU~\citep{MAWU} uses Margin-aware Alignment and Weighted Uniformity to mitigate the bias problem in the dataset, thus achieving better performance.
\section{CONCLUSION}\label{Sec: Conclusion}
In this work, we have discovered that AU-based models suffer from two main problems: weak alignment signals and uniformity affected by extreme values. To overcome these issues, we have introduced two regularizations: center-strengthened alignment and low-variance guided uniformity, and proposed the RAU model. We have conducted extensive experiments on three different real-world datasets with varying sparsities, and the results show that RAU outperforms existing state-of-the-art collaborative filtering methods with a significant performance improvement.

\section*{CRediT authorship contribution statement}
\textbf{Xi Wu}: Conceptualization, Methodology, Software, Writing – original draft.
\textbf{Dan Zhang}: Data curation, Visualization, Writing – review.
\textbf{Chao Zhou}: Conceptualization, Investigation, Validation, Writing – review \& editing.
\textbf{Liangwei Yang}: Formal analysis, Writing – review.
\textbf{Tianyu Lin}: Data curation, Software.
\textbf{Jibing Gong}: Funding acquisition, Resources, Supervision.

\section*{Declaration of competing interest}
The authors declare that they have no known competing financial interests or personal relationships that could have appeared to influence the work reported in this paper.

\section*{Acknowledgment}
This work was supported by Central Government Guidance Local Science and Technology Development Fund Project(246Z0306G), CIPSC-SMP-Zhipu.AI Large Model Cross-Disciplinary Fund, and Innovation Capability Improvement Plan Project of Hebei Province (22567626H).

\section*{Data availability}
Data will be made available on request.

\bibliographystyle{elsarticle-harv} 
\bibliography{main}

\end{document}